# Linear nonsaturating magnetoresistance in kagome superconductor CsV$_3$Sb$_5$ thin flakes

Xinjian Wei[1†], Congkuan Tian[1,2†], Hang Cui[2], Yongkai Li[3,4,5], Shaobo Liu[1,2], Ya Feng[1], Jian Cui[1], Yuanjun Song[1], Zhiwei Wang[3,4,5], Jian-Hao Chen [1,2,6,7*]

[1] Beijing Academy of Quantum Information Sciences, Beijing 100193, China

[2] International Center for Quantum Materials, School of Physics, Peking University, Beijing 100871, China

[3] Centre for Quantum Physics, Key Laboratory of Advanced Optoelectronic Quantum Architecture and Measurement (MOE), School of Physics, Beijing Institute of Technology, Beijing 100081, China

[4] Beijing Key Lab of Nanophotonics and Ultrafifine Optoelectronic Systems, Beijing Institute of Technology, Beijing 100081, China

[5] Material Science Center, Yangtze Delta Region Academy of Beijing Institute of Technology, Jiaxing 314011, China

[6] Key Laboratory for the Physics and Chemistry of Nanodevices, Peking University, Beijing 100871, China

[7] Interdisciplinary Institute of Light-Element Quantum Materials and Research Center for Light-Element Advanced Materials, Peking University, Beijing 100871, China

[†]*These authors contributed equally to this work.*

Corresponding Author
*E-mail: Jian-Hao Chen (chenjianhao@pku.edu.cn)

## Abstract

Linear nonsaturating magnetoresistance (LMR) represents a class of anomalous resistivity response to external magnetic field that has been observed in a variety of materials including but not limited to topological semi-metals, high-$T_c$ superconductors and materials with charge/spin density wave (CDW/SDW) orders. Here we report the observation of LMR in layered kagome superconductor and CDW material CsV$_3$Sb$_5$ thin flakes, as well as the dimensional crossover and temperature ($T$) crossover of such LMR. Specifically, in ultrathin CsV$_3$Sb$_5$ crystals, the magnetoresistance (MR) exhibits a crossover from LMR at low $T$ to quadratic $B$ dependence above the CDW transition temperature; the MR also exhibits a crossover from LMR to sublinear MR for sample thickness at around ~20 nm at low $T$. We discuss several possible origins of the LMR and attribute the effect to two-dimensional (2D) CDW fluctuations. Our results may provide a new perspective for understanding the interactions between competing orders in kagome superconductors.

Keywords: kagome materials, linear magnetoresistance, charge density wave

## 1. Introduction

LMR is an unconventional transport phenomenon that the longitudinal electrical resistivity of the material increases linearly with external magnetic field ($B$). LMR is scientifically interesting[1-6] and potentially useful in applications such as magnetic sensors,[7-9] attracting considerable attention from the materials science community. LMR materials include narrow band-gap disordered semiconductors,[1,7,8,10-13] topological semi-metals with linear dispersion relation,[2,4,14,15] various materials with CDW/SDW orders,[16-18] and high-$T_c$ superconductors.[6] However, an important type of topological materials, e.g., layered materials with kagome lattice, has not been found to exhibit LMR before.

Kagome metals/insulators have a distinctive band structure, simultaneously possessing flat bands, van Hove singularities and topological surface states, together with the potential to form unique spin/charge orders due to their unique lattice geometry.[19,20] Thus, kagome materials provide an ideal arena for studying the interplay among electron-electron correlations, topological nontrivial states and lattice symmetry. Recently, a new family of kagome metal $AV_3Sb_5$ (A = K, Rb, Cs) has attracted great interest with intertwined exotic orders including unconventional CDW, superconductivity, topological nontrivial states and the possibility of additional hidden orders.[20-23] The unconventional CDW order is manifested in magnetic susceptibility, resistivity and specific heat in the temperature range of 70~100 K.[20,21,24] It can also be directly detected by optical spectroscopy,[25,26,27] scanning tunneling microscopy (STM)[22,28,29] as well as angle-resolved photoemission spectroscopy (ARPES).[30-32] According to measurements of magnetization, resistivity, and heat capacity, $CsV_3Sb_5$ and $KV_3Sb_5$ show an onset of superconductivity at 2.5 and 0.93 K respectively.[20,24] For $RbV_3Sb_5$, The onset temperature of superconducting transition in resistance is about 0.92 K and zero resistance temperature sets in about 0.75 K.[33] Unconventional CDW and superconductivity in $AV_3Sb_5$ are shown to coexist and compete with each others,[34-39] and they are both highly tunable by external pressure[36,38,40] and chemical doping[41-43]. To date, the overwhelming majority of the research effort is on bulk $AV_3Sb_5$ crystals, where LMR is not observed. Here, we report the discovery of LMR in $CsV_3Sb_5$ (CVS) thin flakes (<20 nm) at low temperatures (<30 K) as well as the existence of dimensional crossover and temperature crossover of such LMR.

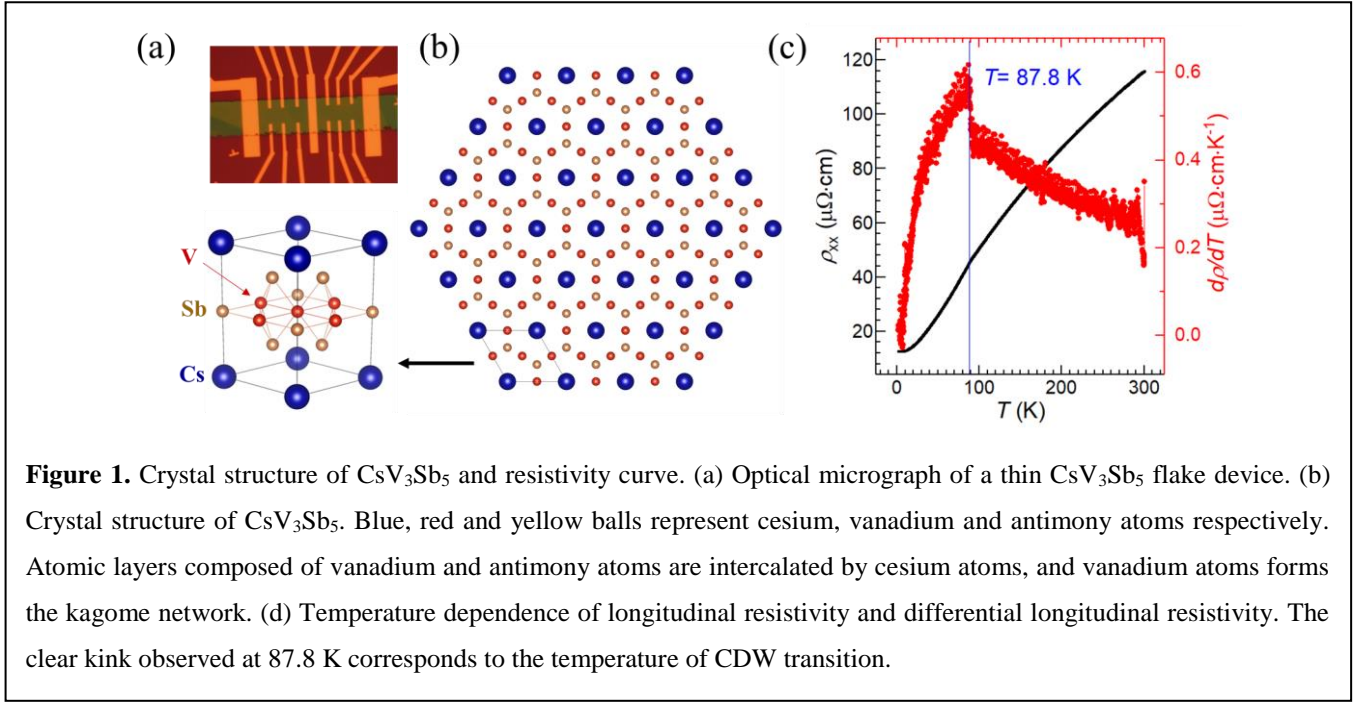

**Figure 1.** Crystal structure of $CsV_3Sb_5$ and resistivity curve. (a) Optical micrograph of a thin $CsV_3Sb_5$ flake device. (b) Crystal structure of $CsV_3Sb_5$. Blue, red and yellow balls represent cesium, vanadium and antimony atoms respectively. Atomic layers composed of vanadium and antimony atoms are intercalated by cesium atoms, and vanadium atoms forms the kagome network. (d) Temperature dependence of longitudinal resistivity and differential longitudinal resistivity. The clear kink observed at 87.8 K corresponds to the temperature of CDW transition.

## 2. Result and Discussion

CVS has an A-A stacked hexagonal lattice structure at room temperature (unit cell of CVS shown in the lower left panel of **Figure 1**(b)). The CVS crystal consists of alternately stacked alkali-metal Cs layers and V-Sb layers along the *c* axis. Figure 1(b) shows the schematic top view of
the crystal structure showing the V and Sb atoms. The V atoms form a 2D kagome framework with in-plane and out-of-plane covalent bonds to Sb atoms; the V-Sb layers are then intercalated by Cs layers with trigonal symmetry. The relatively weak interactions between the V-Sb layers and the Cs layers make CVS easily cleavable. Thin flakes of CVS single crystals are fabricated into Hall bar devices as shown in Figure 1(a). Note that all the device fabrication processes were carried out in an argon glove box or in vacuum, short exposure to air is made only when the device is protected by 160 nm of PMMA layer to prevent sample degradation.

The unconventional CDW in CVS results in clear electrical signal in the longitudinal resistivity $\rho_{xx}$. Figure1 (c) shows the *T* dependence of $\rho_{xx}$ as well as $d\rho_{xx}/dT$ of a CVS thin flake device (sample #3, 12 nm thick). With decreasing temperature, $\rho_{xx}$ decreases monotonically with a kink at around 90 K, corresponding to the CDW transition temperature $T_{CDW}$. The CDW transition is more visible in the $d\rho_{xx}/dT$ vs. *T* curve, where a discontinuity appears in the curve at $T_{CDW}$, which is consistent with previous reports.[29] We extract the precise $T_{CDW}$ ~87.8 K from the $d\rho_{xx}/dT$ vs. *T* curve.

**Figure 2**(a) shows the longitudinal magnetoresistance of sample #3 (12 nm thick) under perpendicular magnetic field and in the temperature range of 2 K to 120 K. The most striking feature in the MR is its nonsaturating linearity with respect to $B$ at low $T$. With increasing $T$, the MR vs. $B$ curves gradually deviate from a linear form to a quadratic form. To quantify such evolution of MR, the MR curves at different $T$ are fitted to a polynomial: $MR=\alpha+\beta B^n$, where $\alpha$, $\beta$ and $n$ are fitting parameters. As shown in Figure 2(b), the polynomial fits to the MR curves very well for all the temperatures explored experimentally. The exponent $n$ has a monotonic dependence on $T$, which is plotted in Figure 2(c). It can be seen that at and above $T_{CDW}$, $n \sim 2$, which means that the MR above the CDW transition temperature are of conventional quadratic dependence on $B$. Below $T_{CDW}$, $n$ reduces dramatically as $T$ decreases, indicating that the reduction of $n$ is closely related to the onset of the CDW phase transition. When $T$ is below 30 K, $n \sim 1$ and remains unchanged, showing that MR vs. $B$ is highly linear and nonsaturating up to 14 T.

To further study this intriguing LMR, we measured the thickness-dependent MR of CVS single crystals. **Figure 3**(a) plots the MR curves for CVS with different thickness at 10 K. In contrast to the relatively thin samples, the MR of thicker crystals becomes sublinear with $B$. The same polynomial $MR=\alpha+\beta B^n$ is fitted to the MR curves, and the fitting parameter $n$ has an abrupt change from 1 to 0.7 for sample thickness larger than 20 nm (Fig 3(b)). Such a thickness-dependent behavior is consistent with a dimensional crossover in the CDW transition in CVS crystals. Indeed, it has been previously reported that $VSe_2$, another layered CDW material, has a crossover in the Fermi surface topology from 3D to 2D at around sample thickness of 20 nm.[44] It is also reported that bulk CVS has 3D CDW,[45-47] while a dimensional crossover of such CDW order from 3D to 2D has been reported via transport measurement without magnetic field.[48] The above evidence suggests that the unusual LMR discovered in CVS thin flakes is likely connected to the 2D nature of the CDW order.

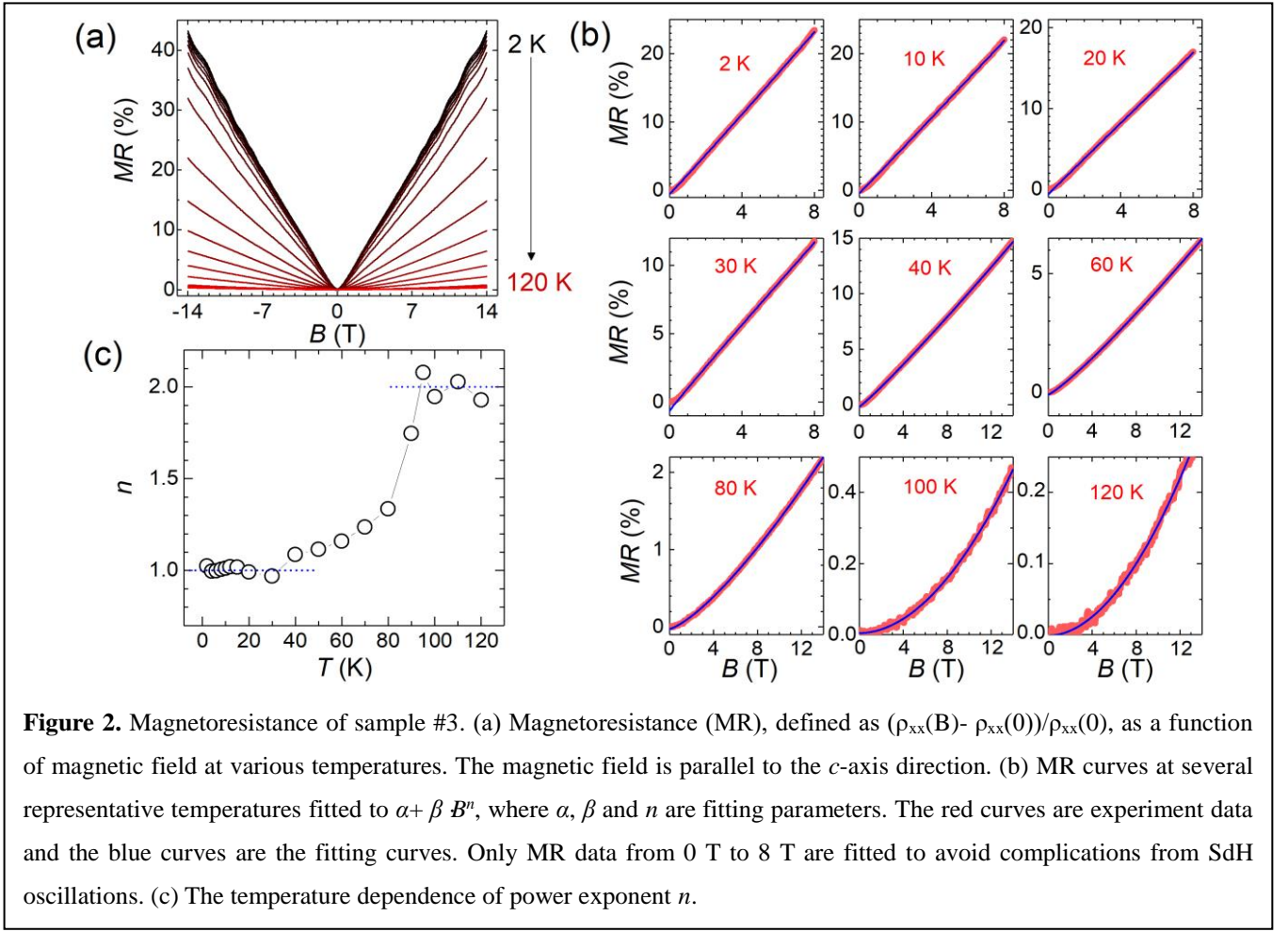

**Figure 2.** Magnetoresistance of sample #3. (a) Magnetoresistance (MR), defined as $(\rho_{xx}(B)-\rho_{xx}(0))/\rho_{xx}(0)$, as a function of magnetic field at various temperatures. The magnetic field is parallel to the $c$-axis direction. (b) MR curves at several representative temperatures fitted to $\alpha+\beta B^n$, where $\alpha$, $\beta$ and $n$ are fitting parameters. The red curves are experiment data and the blue curves are the fitting curves. Only MR data from 0 T to 8 T are fitted to avoid complications from SdH oscillations. (c) The temperature dependence of power exponent $n$.

Now we turn to the discussion on the possible origin of the LMR. Conventionally, magnetic field changes the electron trajectory and reduces effective electron mean path, giving rise to a positive MR which scales quadratically with $B$.[49] However, MR could deviate from the quadratic law due to quantum effects, unusual topological band structures and disorders.[8,10,50] In particular, LMR has been observed in other materials, such as semiconductors with narrow band gap, unconventional superconductors, topological and charge/spin density wave materials.[1,6,15,51-53] Several mechanisms have been proposed to explain LMR. These mechanisms are summarized in the followings and are compared with our experimental findings: 1) The Parish-Littlewood model[8]. This model considers the classical effects of macroscopic disorders and strong inhomogeneity, which lead to the mixing of longitudinal resistance and transverse resistance. With an additional requirement of linear Hall effect, LMR would appear. However, CVS does not have a linear Hall effect both as thin flakes (see **Figure 4**(a)) and as bulk crystals below $T_{CDW}$.[54] In addition, LMR from the Parish-Littlewood model has no obvious temperature dependence and can emerge at high temperature, which is not the case for CVS. Therefore, the Pairth-Littlewood model could not explain our experimental findings. 2) The Abrikosov model.[10] This model considers a gapless semiconductor with a linear dispersion relation and has only one Landau band participating in transport. Although experiments and theoretical calculations have confirmed that bulk CVS has a nontrivial topological band structures with a linear energy spectrum,[20,55] we have not observed strong evidence of nontrivial topological band structures in ultrathin samples from the analysis of SdH oscillation (see supplementary **Figure S2**(c)). On the other hand, LMR in thin CVS sets in at ~0.85 T and goes on beyond 8 T. Considering that two main frequencies of SdH oscillations are found ($F_1$= 31 T and $F_2$= 64 T, see supplementary Figure S2 (a)), it is impossible for only one Landau level to participate in the transport in the magnetic field from 0.85 T to 8 T. Hence Abrikosov's model can't explain our results. 3) The "hot spots" model.[56] This model considers particular areas or points on the Fermi surface (called "hot spots") that scatter charge carriers more than other areas. Under nonzero magnetic field, hot spots provide a scattering rate $1/\tau_{hs}$ which is linearly dependent on $B$, which lead to LMR. The linear $B$ dependence of $1/\tau_{hs}$ can be understood by a simple picture that in the reciprocal space, the frequency of electrons passing through these hot spots (~$1/\tau_{hs}$) is proportional to the cyclotron frequency $\omega_c$, where $\omega_c$ increases linearly with $B$. Thus, $\rho_{xx}$~$1/\tau_{hs}$~$\omega_c$~$B$, and the "hot spots" model may capture the physics of LMR in CVS.

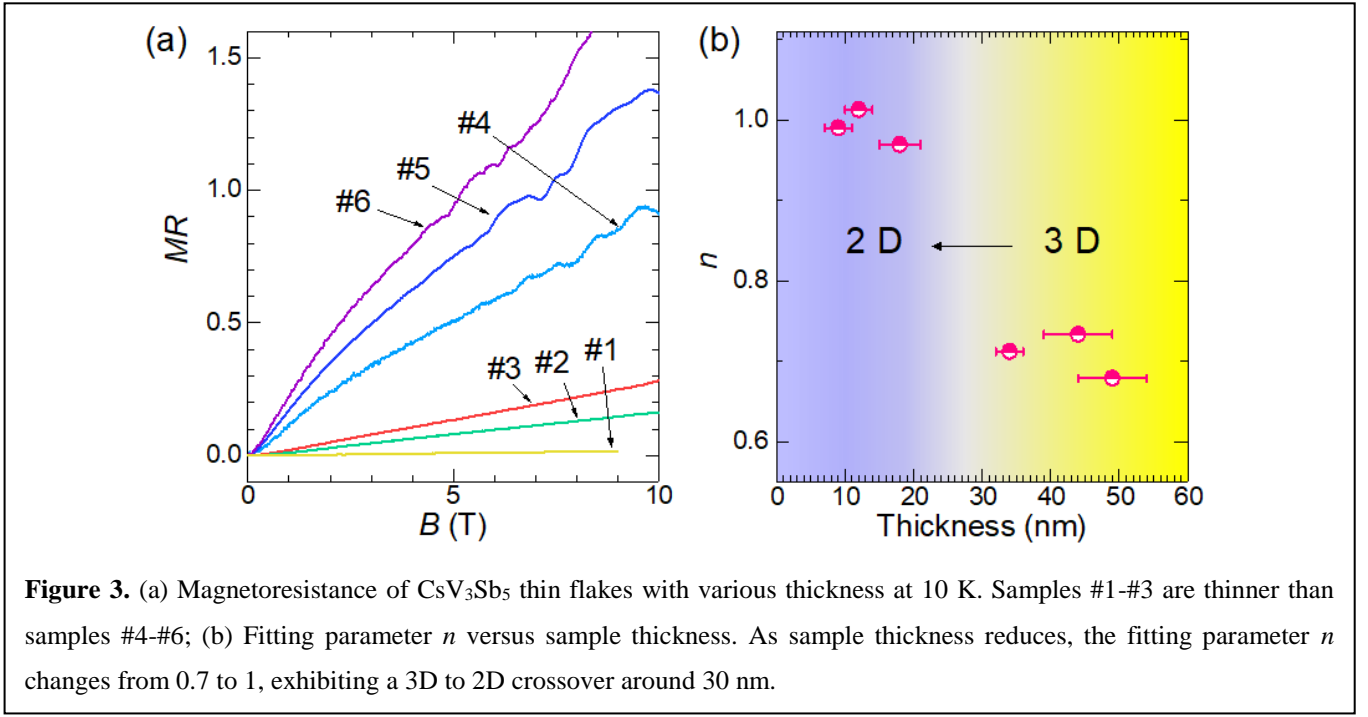

**Figure 3.** (a) Magnetoresistance of $CsV_3Sb_5$ thin flakes with various thickness at 10 K. Samples #1-#3 are thinner than samples #4-#6; (b) Fitting parameter $n$ versus sample thickness. As sample thickness reduces, the fitting parameter $n$ changes from 0.7 to 1, exhibiting a 3D to 2D crossover around 30 nm.

Next, we look into mechanisms to create such hot spots in CVS. First, Young et al. proposed that the Fermi surface close to Brillouin zone boundary can be regarded as hot spots because of intense Umklapp phonon scattering.[56] However, Umklapp scattering strongly depends on temperature, while LMR in thin CVS remains invariant to temperature for $T<30$ K. Second, hot spots could be attributed to magnetic breakdown,[57] which is inconsistent with the fact that LMR in CVS happens at very small magnetic field of 0.8 T and remains nonsaturating up to 14 T. Third, Fermi surface with sharp corners[18] has been proposed to be the source of hot spots. Current ARPES experiment in CVS did show the possibility of Fermi surface with sharp corners, but the resolution is limited. Furthermore, the ARPES experiments are done with bulk CVS crystals that do not show LMR.

With the above discussions, we have narrowed down the possible mechanism to CDW fluctuations[16] that could simultaneously in line with the experimental observation of temperature crossover and dimension crossover in CVS thin flakes. Specifically, 1) the exponent $n$ in $MR=\alpha+ \beta B^n$ has a sharp reduction starting at $T = T_{CDW}$, and 2) LMR appears only in relatively thin films (a dimensional crossover appear at crystal thickness of about 20 nm). The first point shows that LMR in CVS is associated with the onset of CDW. Indeed, recent researches show that the CDW in CVS is driven by the scattering of electrons between neighboring M points.[30] In our experiment, the Fermi surface reconstruction induced by CDW is supported by a deviation from linearity of Hall resistivity $\Delta\rho_{xy}$, such that $\Delta\rho_{xy}$ increases rapidly below $T_{CDW}$ (see Figure 4(b)). Fermi surface reconstruction actually leads to a complex multiband and consequently gives rise to nonlinear $B$ dependent $\rho_{xy}$, which could be a joint effect of classical multiband and nontrivial Berry phase. Here, $\Delta\rho_{xy}(B)$ is extracted from $\rho_{xy}(B)$ by subtracting a linear background obtained from the $\rho_{xy}(B)$ curves above 5 Tesla, and Figure 4(b) plots $\Delta\rho_{xy}(B = 2\ T)$ vs. temperature. Secondly, since reduced dimension is known to increase fluctuations, destroying lattice order,[58] magnetic order[59] and suppressing CDW,[60] LMR in CVS thin flakes could stem from 2D CDW fluctuations. Inspired by the hot spots model, we propose that 2D CDW fluctuation could turn the M points to be hot spots for quasi-particle scattering (see Figure 4(c)) and lead to LMR in CVS. Since the LMR observed in CVS thin flakes is closely related to the formation of CDW as well as the fluctuation of it at reduced dimension, we expect the temperature and thickness dependent MR of other CVS variants to be significantly modified by the nature of their respective CDW correlations.

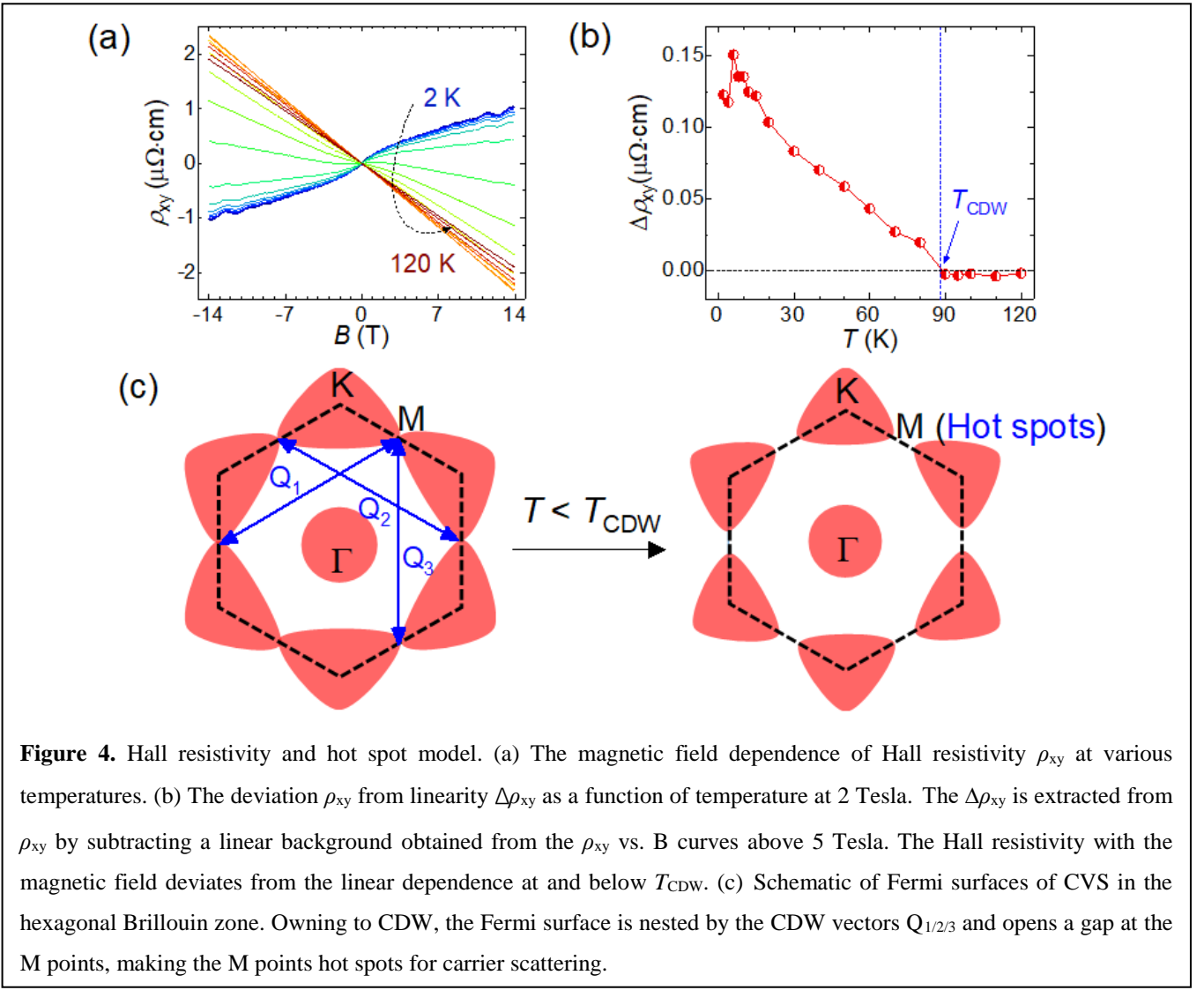

**Figure 4.** Hall resistivity and hot spot model. (a) The magnetic field dependence of Hall resistivity $\rho_{xy}$ at various temperatures. (b) The deviation $\rho_{xy}$ from linearity $\Delta\rho_{xy}$ as a function of temperature at 2 Tesla. The $\Delta\rho_{xy}$ is extracted from $\rho_{xy}$ by subtracting a linear background obtained from the $\rho_{xy}$ vs. B curves above 5 Tesla. The Hall resistivity with the magnetic field deviates from the linear dependence at and below $T_{CDW}$. (c) Schematic of Fermi surfaces of CVS in the hexagonal Brillouin zone. Owning to CDW, the Fermi surface is nested by the CDW vectors $Q_{1/2/3}$ and opens a gap at the M points, making the M points hot spots for carrier scattering.

In thin flakes CVS at finite temperature, electron scattering is likely dominated by 2D CDW fluctuations, impurities and phonons. The total scattering rate, $\tau_{total}^{-1}$, is a sum of the contribution from $\tau_{CDW}^{-1}$ (from 2D CDW fluctuations), $\tau_i^{-1}$ (from impurities) and $\tau_{ph}^{-1}$ (from phonons). Thus $\rho_{xx} = \frac{m}{n_0 e^2}\tau_{total}^{-1} = \frac{m}{n_0 e^2}(\tau_{CDW}^{-1} + \tau_i^{-1} + \tau_{ph}^{-1})$, where $m$ is the electron effective mass, $n_0$ is the carrier density and $e$ is the elemental charge. Above $T_{CDW}$, since there is no CDW order, $\tau_{total}^{-1} = \tau_i^{-1} + \tau_{ph}^{-1} \sim B^2$, leading to the conventional quadratic magnetoresistance. Because 2D CDW fluctuations appear below $T_{CDW}$, according to the hot spots model, $\tau_{CDW}^{-1}$ linearly depends on magnetic field, i.e. $\tau_{CDW}^{-1} \propto B$. According to Matthiessen's rule for incoherent scatterers, MR below $T_{CDW}$ will be proportional to a sum of the first and second power of the magnetic field (thus the exponent $n$ in MR=$\alpha + \beta B^n$ has values between 1 and 2). As temperature is considerably lower than $T_{CDW}$ (i.e., $T<30$ K) and for strong enough magnetic field, $\tau_{CDW}^{-1} \gg \tau_i^{-1} + \tau_{ph}^{-1}$ and $\tau_{total}^{-1} \sim \tau_{CDW}^{-1}$. Therefore, MR becomes linear with magnetic field. This phenomenological theoretical model can explain our results well, and a similar model has been reported to explain LMR in the CDW materials TbTe$_3$ and HoTe$_3$.[16]

Furthermore, superconductivity and LMR coexists in a number of unconventional superconductors, where LMR is considered to be related to specific pairing mechanisms of the superconductor.[6, 51] For example, LMR in cuprate superconductors may be associated with quantum criticality,[6] and LMR in ultrathin superconducting FeSe films is considered to arise from spin fluctuation.[51] Our study of the relation of LMR and CDW in kagome superconductor CVS thus provides a new example the competition and intricate relations of quantum orders in materials.

## 3. Conclusion

In summary, we systematically studied the magnetoresistance and Hall effect of the topological kagome metal CsV$_3$Sb$_5$, and discovered an unusual LMR with both temperature crossover and dimensional crossover effect. In particular, LMR develops for crystals thinner than 20 nm and below 30 K. For $T>T_{CDW}$, MR is quadratic in B, and for crystal thickness larger than 30 nm, LMR is not observed from 2 K to 300 K. The temperature-dependent crossover in LMR is accompanied by the appearance of a Hall anomaly, strongly suggesting that both effects arise from Fermi surface reconstruction caused by the CDW order. The dimensional crossover shows that 2D CDW fluctuation is the most likely cause of this unusual linear magnetoresistance.

## 4. Experimental Section

Single crystal CVS is prepared via the self-flux method.[29] Thin flakes of CVS were exfoliated by Al$_2$O$_3$-assisted exfoliation method[61] onto highly doped silicon wafer with 285 nm SiO$_2$. Then, the electrodes with Hall bar geometry as shown in Figure 1 (a) were prepared by a standard microfabrication process. Electrodes composed by a bilayer of Ti/Au (5 nm/100 nm) were deposited onto the flakes using e-beam evaporation in high vacuum (10$^{-7}$ mbar).

Transport measurements are performed at temperatures between 0.5 K and 300 K with magnetic fields up to 14 T in an Oxford Teslatron cryostat and in a Quantum Design PPMS with a 3He insert. Lock-in amplifiers are used to measure longitudinal resistivity ($\rho_{xx}$) and Hall resistivity ($\rho_{xy}$) at a frequency of 77.77 Hz. Magnetic field is applied parallel to the c-axis direction. The thickness of various samples is measured by Atomic Force Microscopy (AFM).

## Author Contributions

J.-H.C. and X.W. conceived idea and designed the project. Z.W. grew the single crystals. X.W. and C.T. performed the fabrication of devices and the low-temperature measurements with the help of Y.F., J.C and Y.S.. J.-H.C. and X.W. wrote the manuscript. All authors discussed the results and commented on the paper.

## Notes

The authors declare no conflict of interest.

**Acknowledgements**

This project has been supported by the National Basic Research Program of China (Grant Nos. 2019YFA0308402, 2018YFA0305604), the National Natural Science Foundation of China (NSFC Grant Nos. 11934001, 11774010, 11921005), Beijing Municipal Natural Science Foundation (Grant No. JQ20002), the National Natural Science Foundation of China (Grant No. 92065109), National Key R&D Program of China (Grant No. 2020YFA0308800), and Beijing Natural Science Foundation (Grants No. Z210006 and No. Z190006). Z.W. thanks the Analysis & Testing Center at BIT for assistance in facility support.

**References**

[1] J. Hu and T. F. Rosenbaum. 2008 Classical and quantum routes to linear magnetoresistance. Nat. Mater. **7**, 697.

[2] D. X. Qu, Y. S. Hor, J. Xiong, R. J. Cava and N. P. Ong. 2010 Quantum Oscillations and Hall Anomaly of Surface States in the Topological Insulator $Bi_2Te_3$. Science **329**, 821.

[3] C. Shekhar, A. K. Nayak, Y. Sun, M. Schmidt, M. Nicklas, I. Leermakers, U. Zeitler, Y. Skourski, J. Wosnitza, Z. Liu, Y. Chen, W. Schnelle, H. Borrmann, Y. Grin, C. Felser and B. Yan. 2015 Extremely large magnetoresistance and ultrahigh mobility in the topological Weyl semimetal candidate NbP. Nat. Phys. **11**, 645.

[4] T. Liang, Q. Gibson, M. N. Ali, M. Liu, R. J. Cava and N. P. Ong. 2015 Ultrahigh mobility and giant magnetoresistance in the Dirac semimetal $Cd_3As_2$. Nat. Mater. **14**, 280.

[5] T. Khouri, U. Zeitler, C. Reichl, W. Wegscheider, N. E. Hussey, S. Wiedmann and J. C. Maan. 2016 Linear Magnetoresistance in a Quasifree Two-Dimensional Electron Gas in an Ultrahigh Mobility GaAs Quantum Well. Phys. Rev. Lett. **117**, 256601.

[6] P. Giraldo-Gallo, J. A. Galvis, Z. Stegen, K. A. Modic, F. F. Balakirev, J. B. Betts, X. Lian, C. Moir, S. C. Riggs, J. Wu, A. T. Bollinger, X. He, I. Božović, B. J. Ramshaw, R. D. McDonald, G. S. Boebinger and A. Shekhter. 2018 Scale-invariant magnetoresistance in a cuprate superconductor. Science **361**, 479.

[7] A. Husmann, J. B. Betts, G. S. Boebinger, A. Migliori, T. F. Rosenbaum and M.-L.Saboungi. 2002 Megagauss sensors. Nature **417**, 421.

[8] Parish, M., and Littlewood, P. 2003 Non-saturating magnetoresistance in heavily disordered semiconductors. Nature **426**, 162.

[9] W. Niu, Y. Gan, Z. Wu, X. Zhang, Y. Wang, Y. Chen, L. Wang, Y. Xu, L. He, Y. Pu and X. Wang. 2021 Large Linear Magnetoresistance of High-Mobility 2D Electron System at Nonisostructural γ-$Al_2O_3$/$SrTiO_3$ Heterointerfaces. Adv. Mater. Interfaces **8.**

[10] A. A. Abrikosov. 1998 Quantum magnetoresistance. Phys. Rev. B **58**, 2788.

[11] R. Xu, A. Husmann, T. F. Rosenbaum, M.-L. Saboungi, J. E. Enderby and P. B. Littlewood. 1997 Large magnetoresistance in non-magnetic silver chalcogenides. Nature **390**, 57.

[12] Z. Ogorelec, A. Hamzic, and M. Basletic. 1999 On the optimization of the large magnetoresistance of $Ag_2Se$. Europhys. Lett. **46**, 56.

[13] M. Lee, T. F. Rosenbaum, M.-L. Saboungi and H. S. Schnyders. 2002 Band-Gap Tuning and Linear Magnetoresistance in the Silver Chalcogenides. Phys. Rev. Lett. **88**, 066602.

[14] A. Narayanan, M. D. Watson, S. F. Blake, N. Bruyant, L. Drigo, Y. L. Chen, D. Prabhakaran, B. Yan, C. Felser, T. Kong, P. C. Canfield and A. I. Coldea. 2015 Linear Magnetoresistance Caused by Mobility Fluctuations in n-Doped $Cd_3As_2$. Phys. Rev. Lett. **114**, 117201.

[15] M. Novak, S. Sasaki, K. Segawa and Y. Ando. 2015 Large linear magnetoresistance in the Dirac semimetal TlBiSSe. Phys. Rev. B **91**, 041203(R).

[16] A. A. Sinchenko, P. D. Grigoriev, P. Lejay and P. Monceau. 2017 Linear magnetoresistance in the charge density wave state of quasi-two-dimensional rare-earth tritellurides. Phys. Rev. B **96**, 245129.

[17] K. K. Kolincio, M. Roman and T. Klimczuk. 2019 Charge density wave and large nonsaturating magnetoresistance in $YNiC_2$ and $LuNiC_2$. Phys. Rev. B **99**, 205127.

[18] Y. Feng, Y. Wang, D. M. Silevitch, J. Q. Yan, R. Kobayashi, M. Hedo, T. Nakama, Y. Onuki, A. V. Suslov, B. Mihaila, P. B. Littlewood and T. F. Rosenbaum. 2019 Linear magnetoresistance in the low-field limit in density-wave materials. Proc. Natl. Acad. Sci. USA **116**, 11201.

[19]     Simeng Yan, David A. Huse and Steven R. White. 2011 Spin-Liquid Ground State of the S = 1/2 Kagome Heisenberg Antiferromagnet. Science **332**, 1173.

[20]     B. R. Ortiz, S. M. L. Teicher, Y. Hu, J. L. Zuo, P. M. Sarte, E. C. Schueller, A. M. M. Abeykoon, M. J. Krogstad, S. Rosenkranz, R. Osborn, R. Seshadri, L. Balents and J. He, S. D. Wilson. 2020 $CsV_3Sb_5$: A $Z_2$ Topological Kagome Metal with a Superconducting Ground State. Phys. Rev. Lett. **125**, 247002.

[21]     B. R. Ortiz, L. C. Gomes, J. R. Morey, M. Winiarski, M. Bordelon, J. S. Mangum, I. W. H. Oswald, J. A. Rodriguez-Rivera, J. R. Neilson, S. D. Wilson, E. Ertekin, T. M. McQueen and E. S. Toberer. 2019 New kagome prototype materials: discovery of $KV_3Sb_5$, $RbV_3Sb_5$, and $CsV_3Sb_5$. Phys. Rev. Mater. **3**, 094407.

[22]     Y. X. Jiang, J. X. Yin, M. M. Denner, N. Shumiya, B. R. Ortiz, G. Xu, Z. Guguchia, J. He, M. S. Hossain, X. Liu, J. Ruff, L. Kautzsch, S. S. Zhang, G. Chang, I. Belopolski, Q. Zhang, T. A. Cochran, D. Multer, M. Litskevich, Z. J. Cheng, X. P. Yang, Z. Wang, R. Thomale, T. Neupert, S. D. Wilson and M. Z. Hasan. 2021 Unconventional chiral charge order in kagome superconductor $KV_3Sb_5$. Nat. Mater. **20**, 1353.

[23]     Li Yu, C. W., Yuhang Zhang, Mathias Sander, Shunli Ni, Zouyouwei Lu, Sheng Ma, Zhengguo Wang, Zhen Zhao, Hui Chen, Kun Jiang, Yan Zhang, Haitao Yang, Fang Zhou, Xiaoli Dong, Steven L. Johnson, Michael J. Graf, Jiangping Hu, Hong-Jun Gao and Zhongxian Zhao. 2021 Evidence of a hidden flux phase in the topological kagome metal $CsV_3Sb_5$. (arXiv:2107.10714).

[24]     B. R. Ortiz, P. M. Sarte, E. M. Kenney, M. J. Graf, S. M. L. Teicher, R. Seshadri and S. D. Wilson. Superconductivity in the $Z_2$ kagome metal $KV_3Sb_5$. 2021 Phys. Rev. Mater. **5**, 034801.

[25]     Z. X. Wang, Q. Wu, Q. W. Yin, C. S. Gong, Z. J. Tu, T. Lin, Q. M. Liu, L. Y. Shi, S. J. Zhang, D. Wu, H. C. Lei, T. Dong and N. L. Wang. 2021 Unconventional charge density wave and photoinduced lattice symmetry change in the kagome metal $CsV_3Sb_5$ probed by time-resolved spectroscopy. Phys. Rev. B **104**, 165110.

[26]     Qiong Wu, Z. X. W., Q. M. Liu, R. S. Li, S. X. Xu, Q. W. Yin, C. S. Gong, Z. J. Tu, H. C. Lei, T. Dong and N. L. Wang. 2021 Revealing the immediate formation of two-fold rotation symmetry in charge-density-wave state of Kagome superconductor $CsV_3Sb_5$ by optical polarization rotation measurement. (arXiv:2110.11306).

[27]     Xiaoxiang Zhou, Yongkai Li, Xinwei Fan, Jiahao Hao, Yaomin Dai, Zhiwei Wang, Yugui Yao and Hai-Hu Wen. 2021 Origin of charge density wave in the kagome metal $CsV_3Sb_5$ as revealed by optical spectroscopy. Phys. Rev. B **104**, L041101.

[28]     L. Nie, K. Sun, W. Ma, D. Song, L. Zheng, Z. Liang, P. Wu, F. Yu, J. Li, M. Shan, D. Zhao, S. Li, B. Kang, Z. Wu, Y. Zhou, K. Liu, Z. Xiang, J. Ying, Z. Wang and T. Wu, X. Chen. 2022 Charge-density-wave-driven electronic nematicity in a kagome superconductor. Nature **604**, 59.

[29]     Z. Wang, Y.-X. Jiang, J.-X. Yin, Y. Li, G.-Y. Wang, H.-L. Huang, S. Shao, J. Liu, P. Zhu, N. Shumiya, M. S. Hossain, H. Liu, Y. Shi, J. Duan, X. Li, G. Chang, P. Dai, Z. Ye, G. Xu, Y. Wang, H. Zheng, J. Jia, M. Z. Hasan and Y. Yao. 2021 Electronic nature of chiral charge order in the kagome superconductor $CsV_3Sb_5$ Phys. Rev. B **104**, 075148.

[30]     Zhengguo Wang, S. M., Yuhang Zhang, Haitao Yang, Zhen Zhao, Yi Ou, Yu Zhu, Shunli Ni, Zouyouwei Lu, Hui Chen, Kun Jiang, Li Yu, Yan Zhang, Xiaoli Dong, Jiangping Hu, Hong-Jun Gao and Zhongxian Zhao. 2021 Distinctive momentum dependent charge-density-wave gap observed in $CsV_3Sb_5$ superconductor with topological Kagome lattice. (arXiv:2104.05556).

[31]     K. Nakayama, Y. Li, T. Kato, M. Liu, Z. Wang, T. Takahashi, Y. Yao and T. Sato. 2021 Multiple energy scales and anisotropic energy gap in the charge-density-wave phase of the kagome superconductor $CsV_3Sb_5$. Phys. Rev. B **104**, L161112.

[32]     Yang Luo, S. P., Samuel M. L. Teicher, Linwei Huai, Yong Hu, Brenden R. Ortiz, Zhiyuan Wei, Jianchang Shen, Zhipeng Ou, Bingqian Wang, Yu Miao, Mingyao Guo, M. Shi, Stephen D. Wilson, J.-F. He. 2021 Distinct band reconstructions in kagome superconductor $CsV_3Sb_5$. (arXiv:2106.01248).

[33]     Qiangwei Yin, Zhijun Tu, Chunsheng Gong, Yang Fu, Shaohua Yan, Hechang Lei. 2021 Superconductivity and Normal-State Properties of Kagome Metal $RbV_3Sb_5$ Single Crystals. Chin. Phys. Lett. **38**, 037403.

[34]     N. N. Wang, K. Y. Chen, Q. W. Yin, Y. N. N. Ma, B. Y. Pan, X. Yang, X. Y. Ji, S. L. Wu, P. F. Shan, S. X. Xu, Z. J. Tu, C. S. Gong, G. T. Liu, G. Li, Y. Uwatoko, X. L. Dong, H. C. Lei, J. P. Sun, J. G. Cheng. 2021 Competition between charge-density-wave and superconductivity in the kagome metal $RbV_3Sb_5$. Phys. Rev. Research **3**, 043018.

[35]     Y. Song, T. Ying, X. Chen, X. Han, X. Wu, A. P. Schnyder, Y. Huang, J. G. Guo, X. Chen. 2021 Competition of Superconductivity and Charge Density Wave in Selective Oxidized $CsV_3Sb_5$ Thin Flakes. Phys. Rev. Lett. **127**, 237001.

[36]     Z. Zhang, Z. Chen, Y. Zhou, Y. Yuan, S. Wang, J. Wang, H. Yang, C. An, L. Zhang, X. Zhu, Y. Zhou, X. Chen, J. Zhou, Z. Yang. 2021 Pressure-induced reemergence of superconductivity in the topological kagome metal $CsV_3Sb_5$. Phys. Rev. B **103**, 224513.

[37]     F. Du, S. Luo, B. R. Ortiz, Y. Chen, W. Duan, D. Zhang, X. Lu, S. D. Wilson, Y. Song, H. Yuan. 2021 Pressure-induced double superconducting domes and charge instability in the kagome metal $KV_3Sb_5$. Phys. Rev. B **103**, L220504.

[38] F. H. Yu, D. H. Ma, W. Z. Zhuo, S. Q. Liu, X. K. Wen, B. Lei, J. J. Ying, X. H. Chen. 2021 Unusual competition of superconductivity and charge-density-wave state in a compressed topological kagome metal. Nat. Commun. **12**, 3645.

[39] C. Mielke, 3rd, D. Das, J. X. Yin, H. Liu, R. Gupta, Y. X. Jiang, M. Medarde, X. Wu, H. C. Lei, J. Chang, P. Dai, Q. Si, H. Miao, R. Thomale, T. Neupert, Y. Shi, R. Khasanov, M. Z. Hasan, H. Luetkens, Z. Guguchia. 2022 Time-reversal symmetry-breaking charge order in a kagome superconductor. Nature **602**, 245.

[40] K. Y. Chen, N. N. Wang, Q. W. Yin, Y. H. Gu, K. Jiang, Z. J. Tu, C. S. Gong, Y. Uwatoko, J. P. Sun, H. C. Lei, J. P. Hu, J. G. Cheng. 2021 Double Superconducting Dome and Triple Enhancement of Tc in the Kagome Superconductor $CsV_3Sb_5$ under High Pressure. Phys. Rev. Lett. **126**, 247001.

[41] Yongkai Li, Qing Li, Xinwei Fan, Jinjin Liu, Qi Feng, Min Liu, Chunlei Wang, Jia-Xin Yin, Junxi Duan, Xiang Li, Zhiwei Wang, Hai-Hu Wen, Yugui Yao. 2022 Tuning the competition between superconductivity and charge order in the kagome superconductor $Cs(V_{1-x}Nb_x)_3Sb_5$. Phys. Rev. B **105**, L180507.

[42] Yuzki M. Oey, Brenden R. Ortiz, Farnaz Kaboudvand, Jonathan Frassineti, Erick Garcia, Rong Cong, Samuele Sanna, Vesna F. Mitrović, Ram Seshadri, Stephen D. Wilson. 2022 Fermi level tuning and double-dome superconductivity in the kagome metal $CsV_3Sb_{5-x}Sn_x$. Phys. Rev. Mater. **6**, L041801.

[43] Haitao Yang, Yuhang Zhang, Zihao Huang, Zhen Zhao, Jinan Shi, Guojian Qian, Bin Hu, Zouyouwei Lu, Hua Zhang, Chengmin Shen, Xiao Lin, Ziqiang Wang, Stephen J. Pennycook, Hui Chen, Xiaoli Dong, Wu Zhou, Hong-Jun Gao. 2021 Doping and two distinct phases in strong-coupling kagome superconductors. (arXiv:2110.11228).

[44] Árpád Pásztor, Alessandro, Scarfato, Céline Barreteau, Enrico Giannini, Christoph Renner. 2017 Dimensional crossover of the charge density wave transition in thin exfoliated $VSe_2$. 2D Mater. **4**, 041005.

[45] B. R. Ortiz, S. M. L. Teicher, L. Kautzsch, P. M. Sarte, N. Ratcliff, J. Harter, J. P. C. Ruff, R. Seshadri, S. D. Wilson. 2021 Fermi Surface Mapping and the Nature of Charge-Density-Wave Order in the Kagome Superconductor $CsV_3Sb_5$. Phys. Rev. X **11**, 041030.

[46] H. Li, T. T. Zhang, T. Yilmaz, Y. Y. Pai, C. E. Marvinney, A. Said, Q. W. Yin, C. S. Gong, Z. J. Tu, E. Vescovo, C. S. Nelson, R. G. Moore, S. Murakami, H. C. Lei, H. N. Lee, B. J. Lawrie, H. Miao. 2021 Observation of Unconventional Charge Density Wave without Acoustic Phonon Anomaly in Kagome Superconductors $AV_3Sb_5$ (A=Rb, Cs). Phys. Rev. X **11**, 031050.

[47] Z. Liang, X. Hou, F. Zhang, W. Ma, P. Wu, Z. Zhang, F. Yu, J. J. Ying, K. Jiang, L. Shan, Z. Wang, X. H. Chen. 2021 Three-Dimensional Charge Density Wave and Surface-Dependent Vortex-Core States in a Kagome Superconductor $CsV_3Sb_5$. Phys. Rev. X **11**, 031026.

[48] B. Q. Song, X. M. K., W. Xia, Q. W. Yin, C. P. Tu, C. C. Zhao, D. Z. Dai, K. Meng, Z. C. Tao, Z. J. Tu, C. S. Gong, H. C. Lei, Y. F. Guo, X. F. Yang, S. Y. Li. 2021 Competing superconductivity and charge-density wave in Kagome metal $CsV_3Sb_5$: evidence from their evolutions with sample thickness. (arXiv:2105.09248).

[49] A. B. Pippard, 1989 Magnetoresistance in metals, Cambridge university press, New York, USA.

[50] Wang, C. M., and Lei, X. L. 2012 Linear magnetoresistance on the topological surface. Phys. Rev. B **86**, 035442.

[51] Qingyan Wang, Wenhao Zhang, Weiwei Chen, Ying Xing, Yi Sun, Ziqiao Wang, Jia-Wei Mei, Zhengfei Wang, Lili Wang, Xu-Cun Ma, Feng Liu, Qi-Kun Xue, Jian Wang. 2017 Spin fluctuation induced linear magnetoresistance in ultrathin superconducting FeSe films. 2D Mater. **4**, 034004.

[52] Q. Niu, W. C. Yu, K. Y. Yip, Z. L. Lim, H. Kotegawa, E. Matsuoka, H. Sugawara, H. Tou, Y. Yanase, S. K. Goh. 2017 Quasilinear quantum magnetoresistance in pressure-induced nonsymmorphic superconductor chromium arsenide. Nat. Commun. **8**, 15358.

[53] S. Thirupathaiah, Y. Kushnirenko, E. Haubold, A. V. Fedorov, E. D. L. Rienks, T. K. Kim, A. N. Yaresko, C. G. F. Blum, S. Aswartham, B. Büchner, S. V. Borisenko. 2018 Possible origin of linear magnetoresistance: Observation of Dirac surface states in layered $PtBi_2$. Phys. Rev. B **97**, 035133.

[54] F. H. Yu, T. Wu, Z. Y. Wang, B. Lei, W. Z. Zhuo, J. J. Ying, X. H. Chen. 2021 Concurrence of anomalous Hall effect and charge density wave in a superconducting topological kagome metal. Phys. Rev. B **104**, L041103.

[55] Y. Fu, N. Zhao, Z. Chen, Q. Yin, Z. Tu, C. Gong, C. Xi, X. Zhu, Y. Sun, K. Liu, H. Lei. 2021 Quantum Transport Evidence of Topological Band Structures of Kagome Superconductor $CsV_3Sb_5$. Phys. Rev. Lett. **127**, 207002.

[56] R. A. Young. 1968 Influence of Localized Umklapp Scattering on the Galvanomagnetic Properties of Metals. Phys. Rev. **175**, 813.

[57] Naito, M., Tanaka, S. 1982 Galvanomagnetic Effects in the Charge-Density-Wave State of $2H-NbSe_2$ and $2H-TaSe_2$. J. Phys. Soc. Jpn. **51**, 228.

[58] Mermin, N. D. 1968 Crystalline Order in Two Dimensions. Phys. Rev. **176**, 250.

[59] Mermin, N. D., Wagner, H. 1966 Absence of Ferromagnetism or Antiferromagnetism in One- or Two-Dimensional Isotropic Heisenberg Models. Phys. Rev. Lett. **17**, 1133.

[60] Yijun Yu, Fangyuan Yang, Xiufang Lu, Yajun Yan, Yong-Heum Cho, Liguo Ma, Xiaohai Niu, Sejoong Kim, Young-Woo Son, Donglai Feng, Shiyan Li, Sang-Wook Cheong, Xian Hui Chen, Yuanbo Zhang. 2015 Gate-tunable phase transitions in thin flakes of 1T-$TaS_2$. Nat. Nanotechnol. **10**, 270.

[61] Yujun Deng, Yijun Yu, Yichen Song, Jingzhao Zhang, Naizhou Wang, Zeyuan Sun, Yangfan Yi, Yi Zheng Wu, Shiwei Wu, Junyi Zhu, Jing Wang, Xian Hui Chen, Yuanbo Zhang. 2018 Gate-tunable room-temperature ferromagnetism in two-dimensional $Fe_3GeTe_2$. Nature **563**, 94.